\documentclass[prl,twocolumn,twoside,showpacs,floatfix]{revtex4}
\usepackage{epsf,times}
\begin{document}
\title{Fluctuation-driven dynamics of the Internet topology}
\author{K.-I.~Goh, B.~Kahng, and D.~Kim \\} 
\affiliation{\mbox{School of Physics and Center for Theoretical Physics,
Seoul National University, Seoul 151-747, Korea}\\
$($to appear in Phys.\ Rev.\ Lett.\ {\rm (March, 2002)}$)$}
\begin{abstract}
We study the dynamics of the Internet topology based on empirical 
data on the level of the autonomous systems. It is found that the 
fluctuations occurring in the stochastic process of connecting and
disconnecting edges are important features of the Internet dynamics.
The network's overall growth can be described approximately
by a single characteristic degree growth rate $g_{\rm eff} \approx 0.016$ 
and the fluctuation strength $\sigma_{\rm eff} \approx 0.14$,
together with the vertex growth rate $\alpha \approx 0.029$.
A stochastic model which incorporates these values and an adaptation
rule newly introduced reproduces several features of the 
real Internet topology such as the correlations between 
the degrees of different vertices.
\end{abstract}
\maketitle
Recently many studies on complex systems~\cite{nature,science} 
paid attention to complex networks~\cite{rev,porto}.
An interesting feature emerging in such complex systems is 
a power-law behavior in the degree distribution, 
$P_{\rm D}(k) \sim k^{-\gamma}$~\cite{physica,ba}, 
where the degree $k$ is the number of edges incident upon a given 
vertex. Recently, Barab\'asi and Albert (BA)~\cite{ba} introduced an 
evolving network model to illustrate such networks, called the scale-free 
(SF) networks, in which the number of vertices $N$ 
increases linearly with time, and a newly introduced 
vertex is connected to already existing vertices following the 
so-called preferential attachment (PA) rule.

Huberman and Adamic (HA)~\cite{ha} proposed another scenario 
for SF networks. They argued that the fluctuation effect arising 
in the process of connecting and disconnecting edges between 
vertices, is the essential feature
to describe the dynamics of the Internet topology correctly. 
In this model, the total number of vertices $N(t)$ increases 
exponentially with time as  
\begin{equation}
N(t)=N(0)\exp(\alpha t).
\end{equation} 
Next, it is assumed that the degree $k_i$ at a vertex $i$ 
evolves through the multiplicative process
\footnote{
Eq.(2) may be regarded as the PA process with fluctuations:\ The degree
growth under the PA rule can be written as $\partial k_i/\partial t = 
2{\cal L}(t)k_i/\sum_jk_j$, where ${\cal L}(t)$ is the edge creation
rate at time $t$. When the number of edges increases
exponentially with time, ${\cal L}(t)$ and $\sum_jk_j$ have the same
time-dependence, so that ${\cal L}(t)/\sum_jk_j$ is a constant.
When fluctuations are added to the constant, it reduces to Eq.(2).
}, 
\begin{equation}\label{multi}
k_i(t+1)=k_i(t)(1+\zeta_i(t+1)), 
\end{equation}
where $\zeta_i(t)$ is the growth rate of the degree 
$k_i$ at time $t$, which fluctuates from time to time.
Thus, one may divide the growth rate $\zeta_i(t)$ into 
two parts,  
\begin{equation}\label{zeta}
\zeta_i(t)=g_{0,i}+\xi_i(t), 
\end{equation}
where $g_{0,i}$ is the mean value over time, and $\xi_i(t)$ 
the rest part, representing fluctuations \footnote{
When the fluctuation part in $\zeta_i$ is ignored, the mean
growth rate $g_{0,i}$ for each vertex $i$ may be regarded as the 
fitness $\eta_i$ introduced in Ref.\onlinecite{ginestra}, but the fluctuation
effect in $\zeta_i$ was not considered in Ref.\onlinecite{ginestra}.
}.  
$\xi_i(t)$ is assumed to be a white noise satisfying 
$\langle\xi_i(t)\rangle=0$ and 
$\langle\xi_i(t)\xi_j(t')\rangle=\sigma_{0,i}^2\delta_{t,t'}\delta_{i,j}$, 
where $\sigma_{0,i}^2$ is the  variance. 
Here $\langle \cdots\rangle$ is the sample average.
For later convenience, we denote the logarithm of the growth factor as $G_i(t)\equiv \ln(1+\zeta_i(t))$.
Then a simple application of the central limit theorem 
ensures that 
$k_i(t)/k_i(t_0)$, $t_0$ being a reference time, 
follows the log-normal distribution for
sufficiently large $t$. 
To get the degree distribution, one needs to collect all 
contributions from different ages $\tau_i$, growth rates 
$g_{0,i}$, standard deviations $\sigma_{0,i}$ and initial degree $k_i(t_0)$. 
HA first assumed that 
$\zeta_i$ are identically distributed so that $g_{0,i} = g_0$ and 
$\sigma_{0,i} = \sigma_0$ for all $i$. 
Then the conditional 
probability for degree, $P_{\rm D}(k,\tau|~k_0)$, that $k_i=k$ at
time $t=t_0+\tau$, given $k_i=k_0$ at $t=t_0$ is given by
\begin{equation}\label{gaussian}
P_{\rm D}(k, \tau|~k_0)=
{{1}\over {k\sqrt{2\pi \sigma_{\rm eff}^2 \tau}}}
\exp\left\{-{\left(\ln \left({k/k_0}\right)-
g_{\rm eff}\tau\right)^2 \over {2 \sigma_{\rm eff}^2\tau}}\right\},
\end{equation}
where $g_{\rm eff} \equiv \langle G_i(t)\rangle$ and
$\sigma_{\rm eff}^2 \equiv \left\langle (G_i(t) - \langle G_i(t)\rangle)^2\right\rangle$.
$g_{\rm eff}$ and $\sigma_{\rm eff}^2$ are related to $g_0$ and $\sigma_0^2$ as 
$g_{\rm eff} \approx g_0 - \sigma_0^2/2$ and $\sigma_{\rm eff}^2 \approx \sigma_0^2$, respectively~\cite{gardiner}.
Since the density of vertices with age $\tau$ is proportional 
to $\rho(\tau)\sim \exp(-\alpha \tau)$, the degree 
distribution collected over all ages becomes  
\begin{equation}
P_{\rm D}(k)\sim \int d\tau \rho(\tau)P_{\rm D}(k,\tau|~k_0) \sim k^{-\gamma}, 
\end{equation}
where 
\begin{equation}
\gamma=1-\frac{g_{\rm eff}}{\sigma_{\rm eff}^2}+
\frac{\sqrt{g_{\rm eff}^2+2\alpha \sigma_{\rm eff}^2}}{\sigma_{\rm eff}^2}.
\label{gamma}
\end{equation}
Therefore, it is instructive to know the effective values of
$g_{\rm eff}$ and $\sigma_{\rm eff}$ to determine the degree exponent. 

The Internet topology on the level of the autonomous systems (AS) has 
been recorded by the National Laboratory for Applied Network 
Research (NLANR)~\cite{nlanr} since November 1997, which 
enables one to study its evolution.
Here a node represents an AS, which is a unit of router policy in
the Internet, consisting of either a single domain or a group of domains.
Analysis of these data sets has been performed by 
several research groups. Some of the findings are as follows: 
First, the numbers of vertices $N(t)$ and edges 
$L(t)$ increase exponentially with time~\cite{fal3}, and 
$L(t)\sim N(t)^{1+\theta}$ with $\theta>0$, 
showing so-called the accelerated growth~\cite{dm}.
Second, the degree distribution follows a power-law with 
exponent, $\gamma \approx 2.2 \pm 0.1$~\cite{fal3,vespig}. Third, 
there occurs the PA behavior in the evolution 
process~\cite{pa9}. Lastly, there exist certain non-trivial correlations
between the degrees of different vertices~\cite{vespig}. 
Meanwhile, Capocci {\it et al.}~\cite{pietro} 
considered the Internet on the levels of both the Internet
Service Providers (ISPs) 
and hosts, in which the relative frequency of
the creation of the ISPs and hosts determines the degree exponent.
 
In this Letter, we analyze the empirical data of the Internet topology 
from the viewpoint of fluctuation-driven adaptive dynamics 
arising in the process of connecting and disconnecting edges. 
We find that the fluctuation effect is indeed essential to 
the dynamics of the Internet topology. 
We measure the growth rate of vertices $\alpha$ and the effective 
values of $g_{\rm eff}$ and $\sigma_{\rm eff}$, and determine the 
degree exponent 
$\gamma$ using Eq.(\ref{gamma}), which is well compared with the 
directly measured value. 
Moreover, using the measured values, we construct a stochastic model 
following the HA idea. In addition, we include 
in our model an adaptive process that favors rewiring towards vertices
with higher degree.
The network structure constructed in this way reproduces 
the topological features of the real Internet,
such as the correlations between the degrees of different vertices. 
 
\begin{figure}[b]
\centerline{\epsfxsize=7.5cm \epsfbox{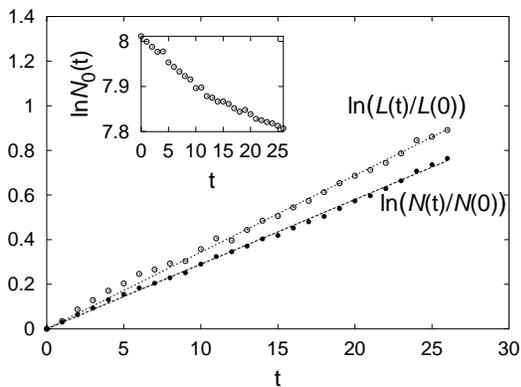}} 
\caption{Semi-logarithmic plot of the total number of 
vertices $N(t)$ ($\bullet$) and edges $L(t)$ ($\circ$) 
as a function of time. The dashed and dotted lines obtained via 
the least-square-fit have slopes 0.029 and 0.034, 
respectively. 
Inset: Semi-logarithmic plot of $N_0(t)$ versus $t$.}
\label{node}
\end{figure}

$Numerical$ $analysis$.\ In our analysis, 
the data are selected monthly, 
so that time is discretized with one month as a unit time step. 
We have analyzed the data from November, 1997 through January, 2000, 
corresponding to $t=0$ and $t=26$, respectively. 
First we examined the total number of vertices $N(t)$ existing 
at time $t$, which grows exponentially with time as 
$N(t)=N(0)\exp(\alpha t)$ with $\alpha \approx 0.029(1)$ 
(see Fig.\ref{node}). The number of edges $L(t)$ also increases 
exponentially with time as $L(t)=L(0)\exp(\beta t)$ with 
$\beta \approx 0.034(2)$ (see Fig.\ref{node}), leading to 
the relation $L(t)=N(t)^{1+\theta}$ with $\theta \approx 0.16(4)$. 
While the total number of vertices increases with time, 
some vertices disappear from the data as time goes on
due to permanent or temporary shutdown of the corresponding AS.
Thus the number of vertices $N_0(t)$
introduced earlier than $t=0$ but still remaining at time $t$, decreases 
with time. Note that the decreasing rate of $N_0(t)$ is considerably 
reduced across $t\approx14$ (see the inset of Fig.\ref{node}).
The hub, the vertex with the largest degree, 
remains identical to the initial one throughout the period we studied.

The average number of degree $k_{\rm new}(t)$ of the vertices 
newly introduced at time $t$ fluctuates in time about a positive mean 
$\langle k_{\rm new}\rangle_t \approx 1.34$ 
with a standard deviation $\sigma_{\rm new}\approx 0.05$,
where $\langle\cdots\rangle_t$ denotes the average 
over a time interval $T$. 
This result suggests that each newly introduced vertex connects 
to only one or two existing vertices, and internal 
links between existing vertices are created actively as time 
goes on. So, the Internet becomes much more interwoven, 
and $L(t)$ grows faster than $N(t)$.

\begin{figure}[b]
\centerline{\epsfxsize=7.5cm \epsfbox{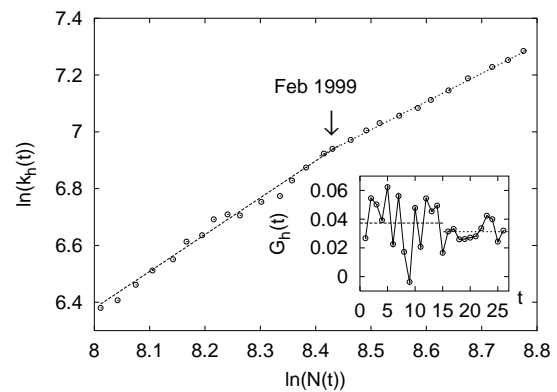}} 
\caption{Plot of the degree of the hub $k_h(t)$ versus 
the number of vertices $N(t)$. Inset: Plot of the growth rate $G_h(t)$
of the degree at the hub versus $t$. The dotted lines 
are the average values for given intervals.} 
\label{hub}
\end{figure}

We consider the dynamics of the degree $k_i(t)$ at each vertex $i$ 
as a function of time. For convenience, we deal with only the vertices 
existing all the time from $t=0$ to $t=26$, and the set 
composed of such vertices is denoted by $S$.  
Thus the degree growth rate of a vertex $i$, 
$G_i(t) = \ln(k_i(t)/k_i(t-1))$ is well defined, because 
$k_i(t)\ne 0$ for any vertex $i \in S$ for all $t$. 
The measured value of $G_i(t)$ fluctuates in time about a finite value. 
Let $g_i \equiv \langle{G_i}\rangle_t$ and $\sigma_i^2 \equiv
\left\langle{(G_i-\langle{G_i}\rangle_t)^2}\right\rangle_t$.
If the dynamics follows that of the HA model, 
a histogram of $g_i$
for many vertices would show the Gaussian distribution with mean
$g_{\rm eff}$ and variance $\sigma_{\rm eff}^2/T$. 
We find that $\{g_i\}$ show some correlations with the degree of the vertex.
In particular, the behavior of 
$G_h(t)$ at the hub is interesting (see the inset of Fig.\ref{hub}).  
It is found that the fluctuation of $G_h(t)$ 
is drastically reduced across $t\approx 15$, February, 1999. 
We obtain $g_h\approx 0.037(20)$ by averaging over the earlier 
period from $t=1$ to $t=15$, while $g_h\approx 0.031(6)$ over the 
later period from $t=16$ to $t=26$. 
Thus, the degree of the hub depends on $N(t)$ as 
$k_h(t) \sim N^{\eta}(t)$, where the exponent $\eta$ 
is related to $g_h$ as $\eta=g_h/\alpha$, exhibiting 
a crossover behavior from $\eta \approx 1.3(1)$ 
to $\eta \approx 1.0(1)$ as directly measured (see Fig.\ref{hub}). 
Note that $\eta = 0.5$ in the BA model.
Despite the apparent correlation of the growth rate
with degree, we do not attempt further analysis
but rather focus on the distribution of $\{g_i\}$ below in accordance with
the HA idea.
Data shown in the inset of Fig.1 and Fig.2 suggest that 
the Internet topology has become much stabilized around $t\approx 15$. 
Therefore, we will use only the data of the later period 
for further discussions. 

\begin{figure}[b]
\centerline{\epsfxsize=7.5cm \epsfbox{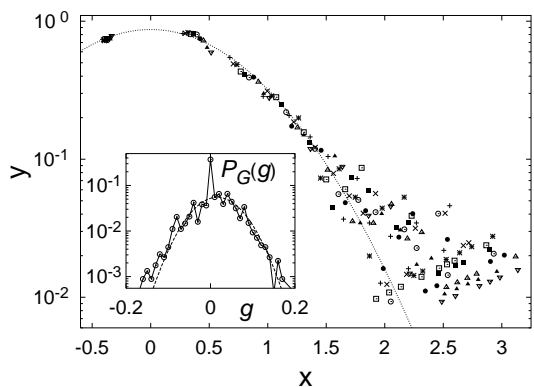}} 
\caption{Plot of $P(k_i(t)/k_i(0))$ versus $k_i(t)/k_i(0)$ for different times 
in terms of $y$ and $x$ defined in the text. The dotted 
line is our best fit of which the peak is located at 
$g_d=0.016$ and the standard deviation is $\sigma_d=0.14$. 
Inset: Plot of the distribution of $g_i = \langle{G_i}\rangle_t$, 
$P_{\rm G}(g)$, versus $g=g_i$ for the vertices in the set $S$. 
The dashed line is our best fit of the central part following the 
Gaussian distribution of which the peak 
is located at ${\bar g}=0.016$ and the standard deviation 
is $\sigma_g=0.04$.}
\label{pk}
\end{figure}

We measured the mean growth rate $g_i$ and the corresponding 
standard deviation $\sigma_i$ defined above for each vertex $i$ ($i \in S$) 
by taking average over the period from $t=16$ to $t=26$ ($T=10$).  
The measured values $\{g_i\}$ ($i \in S$) are distributed as shown 
in the inset of Fig.\ref{pk}. In this inset, an abnormal peak 
is located at $g=0$, which is mostly contributed by the vertices 
whose degree is a few and never changes at all during the 
period we studied. Thus those vertices may be regarded as the ones 
located at dangling ends in the graph, and be ignored for 
the dynamics of the Internet topology.  
In the inset of Fig.\ref{pk}, we fit the data other than the 
$g=0$ peak to the Gaussian form with mean ${\bar g}\approx 0.016(2)$ 
and standard deviation $\sigma_g \approx 0.04$. 
On the other hand, the measured values $\{\sigma_i\}$ are also 
distributed but with small dispersion around the mean 
${\bar \sigma} \approx 0.12(6)$. 
In the HA model, $\bar g$ and $\sigma_g$ would correspond
to $g_{\rm eff}$ and $\sigma_{\rm eff}/\sqrt{T}$, respectively.

It is most likely that $\bar g$ and $\bar{\sigma}$ have a 
distribution among vertices. 
To obtain more accurate degree distribution, Eq.(4) has to be averaged
over those distributions.
Not knowing such details, however, we try to approximate the growth process 
by a single process whose
effective mean growth rate and standard deviation are $g_{\rm eff}$ 
and $\sigma_{\rm eff}$, respectively. For this purpose, we plot in Fig.\ref{pk}
the distribution $P(k_i(t)/k_i(t_0))$ in terms of the scaled variables 
$x$ and $y$ defined as
\begin{equation}
x\equiv {{\ln(k_i(t)/k_i(t_0))-{g_d} (t-t_0)}\over 
{\sqrt {2{{\sigma_d}}^2 (t-t_0)}}},
\end{equation}
and
\begin{equation}
y\equiv P({{k_i(t)}/{k_i(t_0)}})({{k_i(t)}/{k_i(t_0)}})
{\sqrt{2\pi {{\sigma_d}}^2 (t-t_0)}}, 
\end{equation}
using a semi-logarithmic scale, where $g_d$ and $\sigma_d$ are
fitting parameters. We choose $t_0=0$, and the 
data shown in Fig.\ref{pk} are for times $t> 15$. It appears that the data 
for different times collapse onto the curve $\ln y=-x^2$ reasonably well 
for small $x$ with our best choice of $g_d=0.016$ and $\sigma_d=0.14$ 
as shown in Fig.\ref{pk}.
Larger deviations for large $x$ are due to $t$ being finite and are 
caused by the rare statistics of a few nodes whose degree increases
by an anomalously large factor.
The values $g_d$ and $\sigma_d$ are close to 
$\bar g$ and $\bar \sigma$, respectively. 
Also, $\sigma_d/\sqrt{T}\approx 0.044$ is consistent with $\sigma_g$.
We also checked $\sigma_d$ by measuring the variance of 
$P(k_i(t)/k_i(t_0))$ for each time, and plotting them as a function of time. 
The slope of the asymptotic line in the plot 
corresponds to $\sigma_d^2$. Using this method, we also obtain 
$\sigma_d\approx 0.14(1)$, which is in agreement with the one obtained
through the data-collapse method. Thus, the values 
$g_{\rm eff}\approx 0.016$ and $\sigma_{\rm eff}\approx 0.14$ may
be regarded as the effective values representing the growth
process as a single stochastic process.
Applying those values to the formula Eq.(6), we obtain the degree exponent 
$\gamma\approx 2.1$, which is in agreement with the directly 
measured one $\gamma_{\rm AS}\approx 2.2(1)$~\cite{fal3,vespig}. 

$Stochastic$ $model$.\ Using the measured values, 
$\alpha$, $g_{\rm eff}$, and $\sigma_{\rm eff}$, and following the HA idea,
we construct a stochastic model evolving through the following three rules:
(i) Geometrical growth: At time $t$, geometrically increased number of new 
vertices, $\alpha N(t-1)$, are introduced 
in the system, and following the fact $\langle k_{\rm new}\rangle_t 
\approx 1.34$, each of them connects to one or two existing vertices
according to the PA rule. 
(ii) Accelerated growth: Each existing vertex 
increases its degree by the factor $g_0 \approx g_{\rm eff} + \sigma_{\rm eff}^2/2$.
These internal edges are also connected following the PA rule. 
(iii) Fluctuation and adaptation: Each vertex disconnects 
existing edges randomly (resp.\ connects new edges following the PA rule) 
when the noise ($\xi_i(t)$ in Eq.(3)) is chosen to be negative (resp.\ positive). 
This fluctuation has the variance $\sigma_0^2 \approx \sigma_{\rm eff}^2$. 
When connecting, the PA rule is 
applied only within the subset of the existing vertices consisting of 
those having more degree than the one previously disconnected.
This last constraint accounts for the adaptation process in our model. 
Through this adaptation process, the Internet becomes more efficient. 

\begin{figure}[b]
\centerline{\epsfxsize=7.5cm \epsfbox{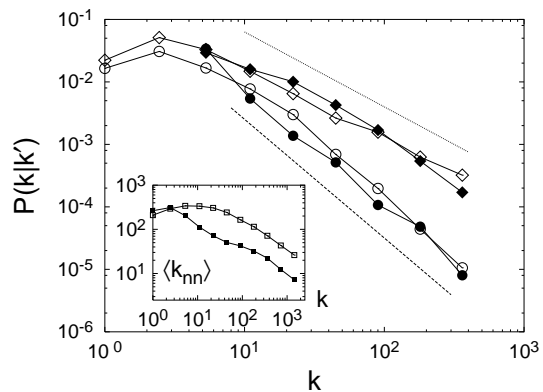}}
\caption{
Plot of the conditional probability $P(k|k')$ for the dangling 
vertices with $k'=1$ (diamonds) and the hub (circles). Data points
with filled (open) symbols are from the real Internet data 
(the model simulations).
The dashed (dotted) line has a slope $-1.9$ ($-1.1$), 
drawn for a guide to the eye.
Inset: Plot of the average degree of nearest neighbors of a vertex
whose degree is $k$, $\langle k_{nn} \rangle= \sum_{k'}k'P(k'|k)$,
as a function of $k$ from the model simulation (open square) and the real
data (filled square).
}
\label{pkk}
\end{figure}

With this stochastic model, we first measure the degree exponent 
to be $\gamma_{\rm model} \approx 2.2$, close to the empirical result 
$\gamma_{\rm AS} \approx 2.2(1)$. 
Second, the clustering coefficient is measured to be 
$C_{\rm model} \approx 0.15(7)$, comparable to the empirical value
$C_{\rm AS} \approx 0.25$ (see also Ref.\onlinecite{vespig}). 
Note that without the adaptation rule, we only get $C\approx
0.01(1)$. Third, we measure the conditional probability, $P(k|k')$,
that the degree of a vertex is $k$ given that it is connected from a vertex
with degree $k'$. 
%It is related to the joint probability, $P(k,k')$,
%that two vertices of degree $k$ and $k'$ are connected,
%via the relation $P(k|k') = P(k,k')/P_{\rm D}(k')$.
For both our model and the real data, it is obtained that 
$P(k|k')\sim k^{-1.1(1)}$ for small $k'$ and
$\sim k^{-1.9(1)}$ for large $k'$ (see Fig.\ref{pkk}).
Note that, for linearly growing networks with the PA rule, it is known that
the probability that vertices of degree $k$ (ancestor) and $k'$ (descendent)
are connected scales as $\sim k^{-(\gamma-1)}k'^{-2}$~\cite{porto,redner}.
%$P(k,k') \sim k^{-(\gamma-1)}k'^{-2}$~\cite{porto,redner}. 
Finally, we measure the average degree of the nearest neighbors 
of a vertex whose degree is $k$, 
$\langle k_{\rm nn}\rangle =\sum_{k'}k' P(k'|k)$, as a function of $k$.
It exhibits a decaying behavior for large $k$,
in agreement with the observation for the real Internet topology,
as shown in the inset of Fig.\ref{pkk} (see also Ref.\onlinecite{vespig}). 
This is in contrast with the $k$-independent behavior occurring
in the BA model.
The adaptive feature in rule (iii) is crucial to reproduce such
detailed agreements between our model and the real data:
While the degree exponent $\gamma_{\rm model} \approx 2.2$ can be 
obtained without the adaptation rule, other results cannot be 
obtained without it. So adaptation as well as fluctuation are 
essential ingredients to describe the real Internet topology correctly. 
Meanwhile, due to the adaptation effect, the network is more 
centralized to the hub, so that it spreads diseases more quickly
via the hub~\cite{epidemic} and becomes more vulnerable to the 
attacks~\cite{attack}.

$Summary$.\ The Internet topology evolves exponentially in the 
number of vertices and edges as time goes on. 
The degree of each vertex also increases 
exponentially with time, but its growth rate fluctuates strongly from time 
to time. 
The effect induced by such fluctuations is essential~\cite{haba}. 
This has not been incorporated properly so far in  
most scale-free network models.
Based on the numerical measurement, we construct a stochastic model
following the HA idea. 
In addition, the adaptation process arising in the evolution 
of edges has been newly taken into account in our model,
through which we can reproduce the correlations between the degrees
of different vertices in the real Internet.

\begin{acknowledgments}
This work is supported by grants No.2000-2-11200-002-3 from the BRP
program of the KOSEF and by the development fund in SNU. 
\end{acknowledgments}

\end{document}